# VISUALIZING PROBABILISTIC PROOF

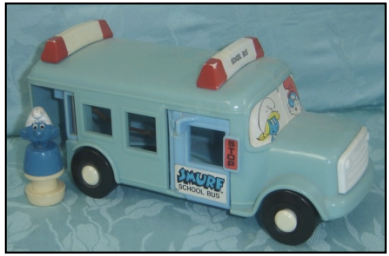

ENRIQUE GUERRA-PUJOL·

ABSTRACT

*The author revisits the Blue Bus Problem, a famous thought-experiment in law involving probabilistic proof, and presents Bayesian solutions to different versions of the blue bus hypothetical. In addition, the author expresses his solutions in standard and visual formats, that is, in terms of probabilities and natural frequencies.*

TABLE OF CONTENTS

---

· B.A., UC Santa Barbara. J.D., Yale Law School. Lecturer in Business Law, Dixon School of Accounting, College of Business Administration, University of Central Florida (UCF). I wish to thank Mark Summers (Barry University School of Law) and Seth J. Chandler (University of Houston Law Center) for reading and criticizing my previous work on probabilistic verdicts and for making me think more deeply about the problem of probabilistic proof. In addition, I wish to thank Krista McCormack and Sydjia Guerra for proofreading previous drafts of this article, and I also thank Sydjia for her love. All remaining errors are mine alone. Credit for the image above: "Kicking the Blues—Smurf Toys," http://kickingtheblues.com/toys.html.

*When the facts change, I change my opinion. What do you do sir?*

—John Maynard Keynes[1]

I. INTRODUCTION: THE PROBLEM OF PROBABILISTIC PROOF

What role should mathematics and mathematical reasoning play in law? Specifically, when should judges and jurors use "probabilistic proof" to infer a defendant's guilt or innocence at trial?

Courts have generally rejected the use of probabilistic proof for identification purposes in both civil and criminal cases.[2] In the eloquent words of one jurist, "Mathematics, a veritable sorcerer in our computerized society, while assisting the trier of fact in the search for truth, must not cast a spell over him."[3] Legal scholars, however, are deeply divided on this question.[4] Broadly speaking, the critics of probabilistic proof point to the dangers of probabilistic methods of adjudication. The proponents of probabilistic reasoning in law, in contrast, emphasize the dangers of not using such probabilistic methods.[5] In this Article, we shall analyze the problem of probabilistic proof through the lens of a thought-experiment, the "Blue Bus Problem," a famous hypothetical or "proof paradox" illustrating the problem of probabilistic proof. Although the blue bus problem is admittedly "an overly simplistic and unusual hypothetical case,"[6] this simple thought-experiment poses fundamental questions about probabilistic proof and the legal process generally. Legal scholars—as well as mathematicians, philosophers, and psychologists—have been debating and discussing the blue bus case and variants of this problem for decades.[7] In summary, solving the blue bus problem is a necessary first step to solving the problem of probabilistic proof.[8]

---

    1. As quoted in Sharon Bertch McGrayne, *The Theory That Would Not Die*. New Haven: Yale University Press (2011).
    2. *See, e.g.*, People v. Collins, 438 P.2d 33 (1968); Smith v. Rapid Transit, Inc., 58 N.E.2d 754 (1945).
    3. *Collins*, 438 P.2d at 33.
    4. For example, compare Laurence Tribe's critique of probabilistic methods in law with Daniel Shaviro's defense of probabilistic proof. *See* Laurence H. Tribe, *Trial by Mathematics: Precision and Ritual in the Legal Process,* 84 HARV. L. REV. 1329 (1971); Daniel Shaviro, *Statistical-Probability Evidence and the Appearance of Justice,* 103 HARV. L. REV. 530 (1989).
    5. For a review of the arguments both for and against probabilistic methods in law, *see* RICHARD LEMPERT & JOSEPH SANDERS, AN INVITATION TO LAW AND SOCIAL SCIENCE: DESERT, DISPUTES, AND DISTRIBUTION (1986).

---

6. Gary L. Wells, *Naked Statistical Evidence of Liability: Is Subjective Probability Enough?*, 62 J. PERS. & SOC. PSYCHOL., no. 5, 1992, at 740.

7. For a small sample of this literature (listed chronologically), with specific reference to the blue bus problem, *see* Edward K. Cheng, *Reconceptualizing the Burden of Proof,* 122 YALE L.J. 1254, 1259, 1269-72 (2013); David Enoch, Levi Spectre, & Talia Fisher, *Statistical Evidence, Sensitivity, and the Legal Value of Knowledge*, 40 PHIL. & PUB. AFF. 197 (2012); S. M. Wexler, *Two Forms of Legal Proof and the So-called Blue Bus Problem*, 6 INT'L ZEITSCHRIFT (IZ) 1 (2010); Gary Wells, *supra* note 6, at 740; Craig R. Callen, *Adjudication and the Appearance of Statistical Evidence,* 65 TUL. L. REV. 457, 494 (1991); Daniel Shaviro, *supra* note 4, at 530–31; Ronald J. Allen, *A Reconceptualization of Civil Trials*, 66 B.U. L. REV. 401, 430 (1986); Lea Brilmayer, *Second-Order Evidence and Bayesian Logic,* 66 B.U. L. REV. 673, 674–76 (1986); Stephen Fienberg, *Gatecrashers, Blue Buses, and the Bayesian Representation of Legal Evidence*, 66 B.U. L. REV. 693 (1986); Stephen Fienberg & Mark Schervish, *The Relevance of Bayesian Inference for the Presentation of Statistical Evidence and for Legal Decisionmaking*, 66 B.U. L. REV. 771, 783–84 (1986); Anne W. Martin, *Comment*, 66 B.U. L. REV. 709, 711 (1986); Charles Nesson, *The Evidence or the Event? On Judicial Proof and the Acceptability of Verdicts*, 98 HARV. L. REV. 1357, 1378 (1985); James Brook, *The Use of Statistical Evidence of Identification in Civil Litigation: Well-Worn Hypotheticals, Real Cases, and Controversy,*" 29 ST. LOUIS U. L.J. 293, 301–03 (1985); Craig R. Callen, *Notes on a Grand Illusion: Some Limits on the Use of Bayesian Theory in Evidence Law*, 57 IND. L.J. 1097 (1982); Michael J. Saks & Robert F. Kidd, *Human Information Processing and Adjudication: Trial by Heuristics*, 15 LAW & SOC'Y REV. 123 (1980); David H. Kaye, *The Paradox of the Gatecrasher and Other Stories*, 1978 ARIZ. ST. L.J. 101, 107–09 (1978). As noted in the text, the source of this famous hypothetical is Tribe, *supra* note 4, at 1340–41.

8. As an aside, another well-known variant of the blue bus problem is "the paradox of the gatecrasher." *See, e.g.*, Cheng, *supra* note 7, at 1270–71; Brilmayer, *supra* note 7, at 674–81, 675–78; Feinberg, *supra* note 7, at 693–97; Fienberg & Schervish, *supra* note 7, at 783–84; Martin, *supra* note 7, at 710–11; Nesson, *infra* note 25, at 522 n.3; Brook, *supra* note 7, at 301–03. *See also* Richard Lempert, *The New Evidence Scholarship: Analyzing the Process of Proof*, 66 B.U. L. REV. 439, 460–62 (1986). The source of the gatecrasher paradox is L. Jonathan Cohen. L. JONATHAN COHEN, THE PROBABLE AND THE PROVABLE § 24, at 74-81 (1977). Yet another (more complex) variant of the blue bus case is the green cab, blue cab problem. *See, e.g.*, Cheng, *supra* note 7, at 1273–74; Michael S. Pardo, *The Nature and Purpose of Evidence Theory*, 66 VAND. L. REV. 547, 588–89 (2013). Judith Jarvis Thomson, *Liability and the Individualized Evidence*, 49 LAW & CONTEMP. PROBS. 199, 199–200 & 203–06 (1986); Maya Bar-Hillel, *The Base-Rate Fallacy in Probability Judgments*, 44 ACTA PSYCHOLOGICA 211, 211–12 (1980); Saks & Kidd, *supra* note 7, at 128, 130. The original sources of the green cab, blue cab problem are the researchers Amos Tversky and Daniel Kahneman. *See* Amos Tversky & Daniel Kahneman, *Evidential Impact of Base Rates*, *in* JUDGMENT UNDER UNCERTAINTY: HEURISTICS AND BIASES 153, 156-57 (Kahneman, et al. eds., 1982). Also see problems 9 & 10 in Amos Tversky & Daniel Kahneman, *Causal Schemas in Judgments Under Uncertainty*, *in* 1 PROGRESS IN SOCIAL PSYCHOLOGY 49, 56, 61 (Martin Fishbein ed., 1980).

My main goal in this paper is to visualize the blue bus problem using Bayesian methods of probabilistic reasoning. Although some scholars have applied Bayesian methods to the blue bus case, none have taken a visual approach to the problem.[9] Ironically, trial attorneys routinely rely on visual forms of evidence (e.g., pictures, objects, scale models, etc.) when presenting their cases to juries.[10] Here, I follow the lead of trial lawyers and present a Bayesian approach to the blue bus problem in visual form using numerical frequencies instead of fractions or percentages.[11] In addition to solving the blue bus problem, my larger aim is to reevaluate the problem of probabilistic proof and make the case for Bayesian methods in civil adjudication generally.[12]

The rest of this paper is organized as follows: Part II reviews the legal case that inspired the blue bus problem, *Smith v. Rapid Transit, Inc.*, and introduces the famous blue bus thought-experiment. Part III presents a Bayesian method for solving the blue bus problem and presents my solution both in the standard Bayesian format (in terms of probabilities or percentages) and in visual form (in terms of natural frequencies or absolute numbers). In brief, the visual representation of the blue bus hypothetical simplifies the problem, making my Bayesian approach easier for judges, jurors, and lawyers to comprehend. Part IV explains why the Bayesian approach to probabilistic proof is a useful method for evaluating all forms of evidence generally. Part V concludes my exposition of a Bayesian visual approach to probabilistic proof.

---

For an overview of these proof paradoxes and others, *see* Mike Redmayne, *Exploring the Proof Paradoxes*, 14 LEGAL THEORY 281 (2008). In this paper, we shall focus solely on the blue bus problem because of its simplicity and pedigree, or in the words of one scholar, "Although the blue bus case is an overly simplistic and unusual hypothetical case, . . . it drives to the heart of the questions of proof, sufficiency, and process." *See* Wells, *supra* note 6, at 740.

9. Some scholars, however, are beginning to work on graphic and visual representations of evidence generally. For a summary of this work, *see* Peter Tiller, *Introduction: Visualizing Evidence and Inference in Legal Settings*, 6 LAW, PROB. & RISK 1 (2007).

10. We thank Orlando I. Martinez-Garcia, personal communication (2001), for bringing this point to our attention. For a diverse sample of visualization techniques in the natural sciences, *see* 2012 *Visualization Challenge*, 339 SCIENCE 509 (Judith Gan & Colin Norman eds., 2013). For the use of visualization techniques in a legal context, *see* Daniel Martin Katz, *Quantitative Legal Prediction*, 62 EMORY L.J. 909 (2013). *See also* Orlando I Martínez-García, *The Person in Law, the Number in Math: Improved Analysis of the Subject as Foundation for a Noveau Régime*, 18 AM. U. J. GENDER SOC. POL'Y & L. 503 (2010).

11. *See generally* Gerd Gigerenzer & Ulrich Hoffrage, *How to Improve Bayesian Reasoning Without Instruction: Frequency Formats*, 102 PSYCHOL. REV., no. 4, 1995, at 684; *see also* Ulrich Hoffrage & Gerd Gigerenzer, *Using Natural Frequencies to Improve Diagnostic Inferences*, 73 ACAD. MED., no. 3, 1998, at 538.

12. We shall consider the use of probabilistic proof in criminal cases in a future paper.

## II. Brief Background

Here, I provide some background into the blue bus hypothetical and the problem of probabilistic proof generally. First, I begin by examining the Federal Rules of Evidence. Next, I review *Smith v. Rapid Transit*, the case that inspired the blue bus hypothetical, and provide some background on the nature of probabilistic proof versus traditional forms of proof. I then turn to the scholarly literature to introduce the blue bus thought-experiment.

### A. The Federal Rules of Evidence

The Federal Rules of Evidence do not address the problem of probabilistic proof directly. One rule broadly states that "[r]elevant evidence is admissible . . ." and that "[i]rrelevant evidence is not admissible."[13] Another rule defines "relevant evidence" as evidence having "any tendency to make a fact more or less probable than it would be without the evidence . . . ."[14] A literal and liberal reading of these rules appears to make room for the use of probabilistic proof at trial, yet in the absence of any particularized or individualized evidence, most courts reject purely probabilistic proof standing alone.[15] Consider the case of *Smith v. Rapid Transit, Inc.* below.

### B. The Original Blue Bus Case (Smith v. Rapid Transit, Inc.)

Since *Smith v. Rapid Transit* provides a simple illustration of the problem of probabilistic proof and is also the source of the famous blue bus problem (see II.C below), I will first explore *Smith*.[16] The plaintiff in this case, Betty Smith, alleged the following facts, as reported in *Smith*, pp. 754–755:

In summary, the plaintiff Betty Smith was driving down Main Street in the City of Winthrop, Massachusetts at about 1:00 a.m. on February 6, 1941.[17] She saw a bus coming toward her at high speed and, to avoid a

---

13. Fed. R. Evid. 402.
14. Fed. R. Evid. 401.
15. *See* cases cited *supra* note 2. Notice that this judicial or common law policy systematically benefits defendants as a class when plaintiffs lack individualized evidence, and in the words of one scholar, "there is no reason why that policy should redound solely to the detriment of plaintiffs." Allen, *supra* note 7, at 430.
16. In a future paper, we shall consider another leading case involving the use of probabilistic reasoning, *People v. Collins,* 438 P.2d 33 (1968).
17. As an aside, notice that the accident occurred in February 1941, while the plaintiff's case was

direct collision with the bus, she swerved to the right and crashed into a parked car, but she did not actually see whose bus ran her off the road. Nevertheless, the plaintiff later discovered that the defendant operated a bus line in the City of Winthrop and had an exclusive license to operate a bus route on Main Street, and according to the defendant's timetable, the defendant's buses were scheduled to travel down Main Street at 12:10 a.m., 12:45 a.m., 1:15 a.m., and 2:15 a.m. Lastly, the plaintiff also discovered that a second bus company had a license to operate a bus line in the City of Winthrop but not on Main Street.

Since this is a civil case, the plaintiff must prove her case by "a preponderance of the evidence."[18] Stated in probabilistic terms, it must be more likely than not that the defendant's bus caused the plaintiff to swerve into the parked car. The problem in the *Smith* case, however, is that the only evidence linking the defendant's bus line to the scene of the accident is probabilistic in nature. That is, since by the plaintiff's own admission she did not actually see which bus was going down Main Street at the time of the accident, the only evidence linking the defendant's bus to the scene of the accident is the defendant's published timetable or schedule.

This problem in turn presents a controversial legal issue. Stated simply, the issue in *Smith* is whether this probabilistic proof alone is enough for the plaintiff to prove her case. Should the plaintiff even be allowed to present her case to the jury? The trial judge ruled that probabilistic proof is not enough as a matter of law and entered a directed verdict in favor of the defendant bus line. The Massachusetts Supreme Court then affirmed the trial court's decision, holding that probabilistic proof, by itself, is not sufficient to prove one's case. The Massachusetts Supreme Court explained its reasoning that "[t]he most that can be said of the evidence in the instant case is that perhaps the mathematical chances somewhat favor the proposition that a bus of the defendant caused the accident. This [is] not enough."[19]

Why is probabilistic proof not enough? If, by the *Smith* Court's own admission, "the mathematical chances somewhat favor the proposition that a bus of the defendant caused the accident," then why is the plaintiff's

---

not finally decided until January 4, 1945. In our view, this lengthy legal delay (lasting nearly four years) alone justifies the consideration of alternatives to traditional litigation. *See, e.g.*, Enrique Guerra-Pujol, *The Turing Test and the Legal Process*, 21 INFO. & COMMC'N. TECH. L. 113 (2012).

18. *See, e.g.*, Wexler, *supra* note 7, at 16. As an aside, a fertile area for further research is the "proof beyond a reasonable doubt" standard in criminal law and the role of Bayesian methods in criminal cases. In a future paper, we shall consider the role of Bayesian reasoning in criminal cases, such as the controversial case of *People v. Collins,* 438 P.2d 33 (1968).

19. *Smith*, 58 N.E.2d at 755.

probabilistic proof not enough to pass the lightweight "preponderance of the evidence" test? The "preponderance of the evidence" standard simply means that the plaintiff's version of events is more likely to be true than the defendant's version of events. Given that that the defendant bus line in *Smith* was the sole bus company authorized to operate on Main Street, then it is highly probable that the defendant's bus was, in reality, the one that caused the plaintiff to swerve and crash.

The *Smith* Court's triumphant but tautological answer to this question was as follows: "While the defendant had the sole franchise for operating a bus line on Main Street, Winthrop, this did not preclude private or chartered buses from using this street; the bus in question could very well have been one operated by someone other than the defendant."[20]

Before proceeding, it is worth noting that this legal ruling, holding that probabilistic proof alone is "not enough," can be restated in Bayesian terms as follows: evidence based on a person's subjective or Bayesian priors alone are never sufficient to pass the "preponderance of the evidence" test. Broadly speaking, note that a "Bayesian prior" refers to one's personal beliefs about the probability of a certain event before one has received or considered any evidence about the likelihood of that event. For example, in Mrs. Smith's case, one might assign an equal probability that the defendant's bus or another bus caused her accident before taking notice of the defendant's timetable or the fact that no other bus company had the legal right to travel on Main Street.[21]

The *Smith* Court's conclusion, however, is circular. Even if the plaintiff herself or another eyewitness were able to identify the particular bus that was going down Main Street at the time of the accident, it is still possible, as the *Smith* Court stated, that "the bus in question could very well have been one operated by someone other than the defendant."[22] After all, the plaintiff's identification testimony would no doubt be self-serving and biased. Further, her vision could have been cloudy, her recollection imperfect.[23] In sum, even if the plaintiff or a neutral witness had seen the actual bus that caused Mrs. Smith to swerve, the evidence would still remain probabilistic in nature, because even if the plaintiff were to offer individualized or particularized proof (as opposed to mere probabilistic proof), it would still be possible that the witness could be mistaken, that "the bus in question could very well have been one operated by someone other than the defendant"![24]

---

20. *Id.*
21. For a readable overview of the Bayesian approach to probability, see NATE SILVER, THE SIGNAL AND THE NOISE: WHY MOST PREDICTIONS FAIL BUT SOME DON'T 232-261 (2012).
22. *Smith*, 58 N.E.2d at 755.
23. For a comprehensive overview of cognitive biases in the adjudication process, *See generally*, Saks & Kidd, *supra* note 7. Since the unreliability of eyewitness testimony is by now well-known, this is why we insist that all forms of evidence, even individualized or particularized evidence, are probabilistic in nature.

Nevertheless, despite its circular reasoning, the *Smith* case poses an important question about the role of probabilistic proof in law.

*C. The Blue Bus Thought-Experiment*

We now turn our attention to the blue bus problem, a famous thought-experiment in evidence law.[25] Broadly speaking, thought-experiments like the blue bus problem are useful because they reveal anomalies, help clarify our thinking, and lead to new insights. Professor Laurence Tribe, the original source of this thought-experiment, presents the problem as follows:

> Consider next [a case] in which the identity of the responsible agent is in doubt. Plaintiff is negligently run down by a blue bus. The question is whether the bus belonged to the defendant. Plaintiff is prepared to prove that defendant operates four-fifths of all the blue buses in town. What effect, if any, should such proof be given?"[26]

Professor Tribe's hypothetical has generated extensive commentary among scholars (see references in footnote 8 above) and has been retold many times over the years. For example, Professor Charles Nesson

---

24. For a detailed and well-reasoned critique of the standard arguments against the use of probabilistic evidence, *see* Brilmayer, *supra* note 7, at 674–81. For brevity, we shall not cover this highly technical terrain in this paper, for our main point is that all evidence is ultimately probabilistic.

25. The blue bus hypothetical has many names. Ed Cheng, for example, refers to "the Blue Bus problem famous in statistical proof circles." Cheng, *supra* note 7, at 1259. Similarly, Steve Wexler, Lea Brilmayer, and Anne Martin also refer to "the blue bus problem." Wexler, *supra* note 7, at 2; Brilmayer, *supra* note 7, at 674; and Martin, *supra* note 7, at 711. For their part, Gary Wells and Charles Nesson refer to "the blue bus case," while Craig Callen refers to "the popular hypothetical from the Smith case." Wells, *supra* note 6, at 740; Nesson, *supra* note 7, at 1378; Callen, *supra* note 7, at 494. In a different paper, Professor Nesson refers to "the famous blue bus hypothetical." *See* Charles Nesson, *Agent Orange Meets the Blue Bus: Factfinding at the Frontier of Knowledge*, 66 B.U. L. REV. 521, 522 (1986). Stephen Feinberg refers to "Laurence Tribe's 'case of the blue bus'" as a "classic hypothetical[ ]." Feinberg, *supra* note 7, at 693. Likewise, Craig Callen refers to "Professor Tribe's blue bus hypothetical . . . ." Callen, *Notes on a Grand Illusion*, *supra* note 7, at 34.

26. Tribe, *supra* note 4, at 1340–41. Although the primary source for the blue bus hypothetical is Tribe's 1971 paper, in a footnote Tribe acknowledges that *Smith v. Rapid Transit* is "the actual case on which this famous chestnut is based." *Id.* at 1341 n.37 Tribe's reference to "this famous chestnut" appears to imply that the blue bus hypothetical was already part of a well-established oral tradition among generations of law professors and law students. For another famous example of "a fairly well-defined oral tradition," this one in economics, see R. H. Coase, *The Problem of Social Cost*, 3 J. L. & ECON. 1, 39–42 (1960) (analysis of factory smoke).

describes the blue bus problem as follows:

> While driving late at night on a dark, two-lane road, a person confronts an oncoming bus speeding down the center line of the road in the opposite direction. In the glare of the headlights, the person sees that the vehicle is a bus, but he cannot otherwise identify it. He swerves to avoid a collision, and his car hits a tree. The bus speeds past without stopping. The injured person later sues the Blue Bus Company. He proves, in addition to the facts stated above, that the Blue Bus Company owns and operates 80% of the buses that run on the road where the accident occurred. Can he win?[27]

For her part, Professor Lea Brilmayer presents the blue bus problem as follows:

> In the blue bus problem, the plaintiff has uncontroverted eyewitness testimony demonstrating that he was run over by a blue bus. However, he does not have a license plate number or any other evidence tending to show which blue bus it was. The defendant in the hypothetical owns four-fifths of the blue buses in town.[28]

Similarly, Professor Daniel Shaviro restates the problem as follows:

> Consider the well-known hypothetical case in which a plaintiff is hit by a bus that she knows is blue but cannot otherwise describe. She establishes that the defendant bus company operates eighty percent of all the blue buses in the area. In the absence of other relevant evidence, this evidence establishes an eighty percent chance that the bus company is liable. Nonetheless, the plaintiff is denied recovery by directed verdict.[29]

Most recently, Professor Edward Cheng presents the problem as follows:

> Recall the facts of *Smith v. Rapid Transit, Inc.*, as famously adapted by Laurence Tribe and Charles Nesson. The plaintiff is driving along a two-lane undivided country road on a dark night when she is faced with the oncoming lights of a bus traveling along the median. To avoid an accident, the plaintiff swerves, causing her car to end up in a roadside ditch. Because of the emergency, the plaintiff is unable to observe anything except that the bus was blue. The plaintiff presents this testimony, along with evidence that the defendant, the Blue Bus Company, operates 80 percent of the blue buses in town. The defense concedes both facts and presents no additional evidence. Is the plaintiff entitled to recover?[30]

---

27. Nesson, *supra* note 7, at 1378–79.
28. Brilmayer, *supra* note 7, at 674.
29. Shaviro, *supra* note 4, at 530–31 (citations omitted). Shaviro's statement of the blue bus problem also appears in Ronald J. Allen, *On the Significance of Batting Averages and Strikeout Totals: A Clarification of the Naked Statistical Evidence Debate, the Meaning of Evidence, and the Requirement of Proof beyond a Reasonable Doubt*, 65 TUL. L. REV. 1093, 1097–98 (1991).

In sum, as this small sample shows, the blue bus problem has been stated and restated in different ways over the years. Nevertheless, all these versions of the blue bus problem share the same information content. Specifically, these various versions of the blue bus problem provide us with two pieces of information:

1. The plaintiff was injured by a blue bus.

2. The defendant operates four-fifths or 80% of all the blue buses in town.

Thus, since all of these statements of the blue bus problem share the same information content, we shall refer to them as the "standard version" of the blue bus problem

Given this information content, the standard version of the blue bus problem is a useful thought-experiment. In the words of the historian of science Thomas Kuhn, for example, a thought-experiment, or the close analysis of an "imagined situation," can confront its audience with unanticipated consequences of their normal conceptual operations.[31] This is precisely why the blue bus hypothetical is useful and illuminating, for it produces a clear anomaly or contradiction. Specifically, when confronted with the blue bus problem, most people are reluctant to impose civil liability on the defendant in this case, even though there is a substantial statistical probability that it was the defendant's blue bus that caused the plaintiff's injuries.[32] Thus, the blue bus problem presents a puzzle: given such a high level of probability of guilt, why does plaintiff's case in this hypothetical not pass the "preponderance of the evidence" test?

---

30. Cheng, *supra* note 7, at 1269 (citations omitted).
31. THOMAS S. KUHN, *A Function for Thought Experiments*, in THE ESSENTIAL TENSION: SELECTED STUDIES IN SCIENTIFIC TRADITION AND CHANGE 240, 252 (1977). By way of example, in his essay on thought experiments Kuhn explains how Galileo's simple thought experiment on falling objects transformed the classical or Aristotelian conception of speed. *Id.* at 249.
32. This is not mere idle speculation or armchair theorizing. *See* Wells, *supra* note 6, for experimental confirmation of this result.

One possible reason for this anomaly, however, is simply the low information content of the blue bus problem. That is, in Bayesian terms, the standard version of the blue bus problem provides information only about true positives or "hits": the fraction of buses that are blue. It does not, however, provide any information about false positives or "false alarms"—the number of buses that are blue but that are not owned by the defendant—nor are we given any information about the total number of buses in town (not just blue buses, but all buses). Given this information gap, it is no wonder why so many legal scholars[33] have rejected the use of probabilistic proof. The standard version of the blue bus problem does not provide us with sufficient information to draw a Bayesian inference about the strength of the evidence in the blue bus case. We are not the first to make this observation,[34] but in part three of the paper below, we are the first to consider different versions of the blue bus problem and to present our Bayesian model of the blue bus problem in visual form.

Putting the issue of information content to one side, the standard version of the blue bus problem also presents a larger question about the role of probabilistic proof or "naked statistical evidence" in law. Legal scholars are sharply divided on this question, offering a wide variety of reasons in support or against the use of probabilistic proof. In summary, the arguments against probabilistic proof often focus on moral intuitions, on the qualitative or non-quantitative aspects of evidence, and on the public perception of the legal system, or in the words of one academic, the "danger in being mesmerized by precision and formalization."[35]

Yet the critics of probabilistic proof all miss an important point: all evidence is ultimately probabilistic in nature. Harvard Law Professor Charles Nesson, for example, has argued that jury verdicts based solely on probabilistic evidence would undermine the public legitimacy of the legal system.[36] In the words of Professor Nesson, "What is crucial in this situation is that the public cannot view whatever statement the factfinder makes as anything other than a bet based on the evidence."[37] This reasoning is just as unpersuasive, and equally as circular, as the court's reasoning in the *Smith* case. After all, if all evidence (statistical or not) is probabilistic in nature, then all jury verdicts are "anything other than a bet on the evidence." Professor Tribe too concurs with the outcome in *Smith*: "cases like *Smith* are entirely sensible if understood instead as insisting on the presentation of *some* non-statistical and 'individualized' proof of identity before compelling a party to pay damages . . . ."[38] Nevertheless, this argument is weak and unpersuasive for the same reason that Nesson's argument fails. In sum, the problem with Tribe's reasoning is that all forms of evidence, even non-statistical evidence and "individualized" proof, are ultimately probabilistic in nature.

---

33. *See, e.g.*, Tribe, *supra* note 4; Nesson, *supra* note 25.
34. *See, e.g.*, Wells, *supra* note 6, at 752 ("there is no other evidence [in the blue bus case] to combine with the base rates."); Fienberg & Schervish, *supra* note 7, at 783 ("the only evidence [in the blue bus case] is a summary statistic"); Martin, *supra* note 7, at 712 (the blue bus problem "do[es] not provide sufficient information to apply Bayes's Theorem").
35. Brilmayer, *supra* note 7, at 691.
36. *See* Nesson, *supra* note 7, at 1378–79.
37. *Id.* at 1379.

In contrast, the arguments in favor of probabilistic evidence generally emphasize the goal of accuracy in adjudication and question the distinction between individualized or particularized evidence and "probabilistic evidence" or "naked statistical evidence." The proponents of probabilistic proof, however, have generally been unable to solve the blue bus problem. Furthermore, their arguments are generally complex, hard to follow, and impractical in a traditional math-phobic legal setting. For example, Stephen Fienberg and Mark Schervish, two leading proponents of Bayesian reasoning in law, have themselves conceded that "implementation of [Bayesian methods], especially in the legal setting, is fraught with difficulties."[39]

In the remainder of this Article, I shall take a different approach to the blue bus problem. Specifically, we shall visualize the blue bus problem from a Bayesian perspective. In sum, a Bayesian approach to the blue bus case allows one to avoid circular reasoning, solves the problem of probabilistic proof by identifying the missing information one needs in order to make Bayesian inferences, and most importantly, provides an accurate alternative to traditional adjudication. In addition, the visual representation of the blue bus hypothetical simplifies the blue bus problem, making our Bayesian reasoning easy to perform and making our Bayesian solutions easy to understand.

### III. A VISUAL APPROACH TO THE BLUE BUS PROBLEM

Here, we present a Bayesian model of Professor Tribe's famous blue bus hypothetical. In addition, we present our model in two formats: (i) in the standard Bayesian format (in terms of probabilities or percentages) as well as (ii) in visual form (in terms of natural frequencies or absolute numbers). Before presenting our model, however, we must mention that the blue bus problem (as stated by Tribe and others) does not have a single solution. That is, to apply Bayesian reasoning to the blue bus problem or any other probabilistic proof problem, we need three pieces of information: (i) the percentage of city buses that run on Main Street (the base rate), (ii) the percentage of buses that run on Main Street that are blue (hits), and (iii) the percentage of blue buses that do not run on Main Street (false alarms or misses).[40] Since the standard version of the blue bus problem (as summarized in Part II.C. above) does not provide us with information about items (i) and (iii), we shall consider different versions of the blue bus problem in Parts 3.A to 3.G below; we solve a variety of blue bus problems with different values for items (i), (ii) and (iii).

---

38. Tribe, *supra* note 4, at 1341 n.37.
39. Fienberg & Schervish, *supra* note 7, at 794.

## A. Example #1 (Standard Probability Format)

As we noted above, we need three pieces of information to apply Bayesian methods to the blue bus problem:

(i) the total number of buses in the population of city buses, or base rate information;

(ii) the number of "hits" or the conditional probability that a bus is blue, given that it runs on Main Street, expressed as "p(blue|Main)" in standard Bayesian notation; and,

(iii) the number of "false alarms" or the conditional probability that a bus is blue, but does not run on Main Street, expressed as "p(blue|not Main)" or p(blue|~Main)" in standard Bayesian notation.

Since the standard version of the blue bus problem (as summarized in Part III.A above) provides information about item (ii) but is silent about items (i) and (iii), for our first version of the blue bus problem we shall assume the following priors:

- 40% of all buses run through Main Street, 60% do not;
- 80% of the buses running down Main Street are blue;[41] and,
- 10% of the buses that do not run on Main Street are blue.

---

40. Christoph Engel & Gerd Gigerenzer, *Law and Heuristics: An Interdisciplinary Venture, in* HEURISTICS AND THE LAW 1, 10 (Christoph Engel & Gerd Gigerenzer eds., 2006) (using the terms "false alarms" and "misses"). *See also* Enrique Guerra-Pujol, *supra* note 17, at 115.

41. This assumption is based on the standard version of the blue bus problem as summarized in Part II.C above. *See also* Tribe, *supra* note 4, at 1340–41.

Given these pieces of information, what is the posterior probability that a bus runs on Main Street, given that it is blue?

Before proceeding, notice that this question is precisely what our Bayesian model of the blue bus problem is designed to test. That is, our Bayesian model of the blue bus problem is designed to test or measure the strength of the plaintiff's hypothesis that it was a blue bus that caused her to swerve. Notice, too, that this question states the blue bus problem in a different way. We are not asking, "What is the probability that the bus in the plaintiff's case was blue?" Instead, we are asking a different question, "What is the probability that a bus will be on Main Street, given that it is a blue bus?"

Now, to perform this critical test, we begin by asking, what do we know? What pieces of information do we have?

First, we already know that 40% of all city buses run through Main Street. This is our "prior probability" and can be stated formally as follows:

$$p(Main) = 0.4$$

Next, we also know that 80% of the buses that run on Main Street are blue. This is our "first conditional probability" and can be stated formally as follows:

$$p(blue|Main) = 0.8$$

In plain language, the probability that a certain bus is blue, given that it runs on Main Street, is 80% or 0.8. Also, notice that this assumption is based on Tribe's famous blue bus problem.[42]

Lastly, we know that 10% of the buses that do not run on Main Street are blue. This is our "second conditional probability" and can also be stated formally as follows:

$$p(blue|\sim Main) = 0.1$$

In plain language, the probability that a certain bus is blue, given that it does not run on Main Street, is 10% or 0.1.

Given these pieces of information, we want to test the plaintiff's blue bus hypothesis by solving for:

$$p(Main|blue) = ?$$

Or, stated formally, given our priors $p(Main)$, $p(blue|Main)$, and $p(blue|\sim Main)$, what is the revised or posterior probability that a certain bus runs on Main Street, given that it is blue? In plain language, what is

---

42. *See id.*

the strength of the plaintiff's hypothesis that a blue bus caused her to swerve?

In brief, we can test plaintiff's blue bus hypothesis using Bayesian reasoning as follows:

1. First, since 40% of city buses run on Main Street, and since 80% of those buses that travel down Main Street are blue, then 32% of all city buses are blue and run through Main Street [0.4 * 0.8 = 0.32];

2. Next, since 60% of the buses do not run down Main Street, and since only 10% of those buses are blue, then 6% of all city buses are blue and do not run through Main Street [0.6 * 0.1 = 0.06];

3. Therefore, since 32% + 6% = 38%, then a total of 38% of the buses that run down Main Street are blue;

4. Lastly, since 0.32 divided by 0.38 = 0.842, this means that there is an 84.2% posterior probability that a certain bus runs on Main Street, *given that it is blue*.

In other words, it is very likely that the bus on Main Street that caused the plaintiff to swerve was a blue bus. Specifically, our conditional probabilities in this example caused us to revise or update our prior probability from 40% to over 84%. Our Bayesian approach has tested the plaintiff's hypothesis, and the plaintiff has passed this test with flying colors.

But why should this Bayesian probability be relevant at trial? That is, after all, the larger question posed by cases like *Smith v. Rapid Transit* and by the blue bus problem in particular.

Recall that the injured plaintiff in the blue bus problem knows where the accident occurred (on Main Street) but does not know the color of the bus that caused her to swerve. *But the fact that the plaintiff knows where the accident occurs tells us something about the probable color of the bus that caused her to swerve!* In brief, this is why Bayesian reasoning is so helpful in solving the blue bus problem.

Example #1 (Visual Format)

Now, we present the above example visually in a "frequency tree" format[43] as follows:

---

43. Gigerenzer & Hoffrage, *How to Improve Bayesian Reasoning*, supra note 11.

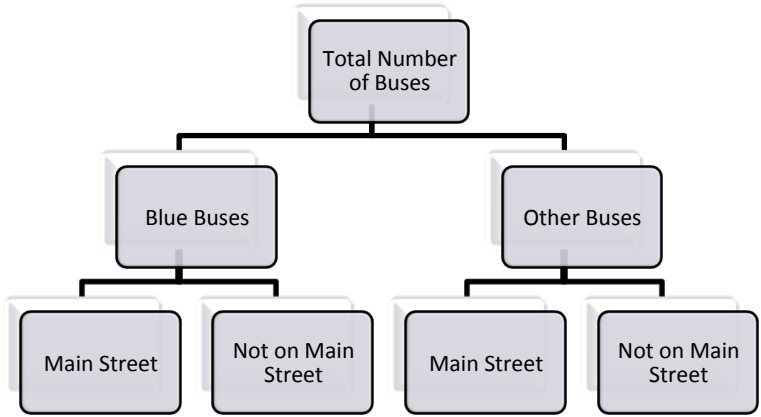

With this visual format as our guide, we now present our first version of the blue bus problem (Example #1) using absolute numbers in place of abstract percentages as follows:

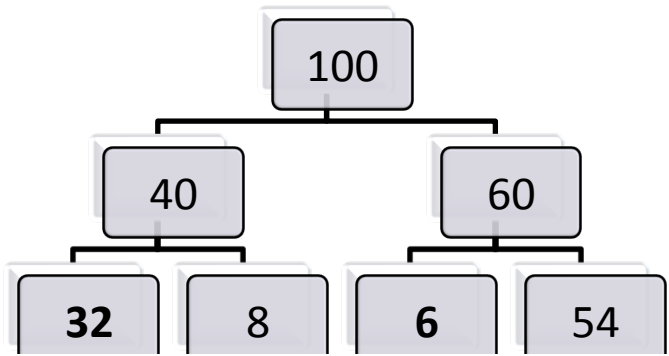

In summary, this frequency tree consists of three rows of information. The top row of the tree tells us the total number of buses in the population of city buses. [For simplicity, we assume there are 100 buses. This way, the absolute numbers in our frequency tree will correspond one-to-one with the probabilities/percentages in the standard probability format above.] The middle row provides the relevant base rate information in absolute numbers, telling us (in this example) that 40 of the 100 buses in town run on Main Street and that the remaining 60 buses do not run through Main Street. In Bayesian terms, the absolute number 40 is equivalent to p(Main) above, or the prior probability that a bus runs on Main Street.

Lastly, the bottom row of the tree contains the relevant conditional probabilities, that is, the critical information regarding "hits," "misses," and "false alarms" that will allow us to perform a Bayesian test to measure the strength of plaintiff's evidence. Before proceeding, notice that the most important pieces of information appear in the first and third branches of the tree, moving from left to right.

Specifically, the first branch on the bottom of the tree (on the left) tells us the number of "hits" in our example, that is, the absolute number of blue buses that run on Main Street. In Bayesian terms, the first branch is equivalent to p(blue|Main), or the conditional probability that a bus is blue given that it runs down Main Street. In this example, 32 of the 40 buses that run on Main Street are blue buses.

By contrast, the third branch on the bottom of the tree (going from left to right) tells us the absolute number of "false alarms," the number of blue buses that do not run on Main Street. In this particular example, there are six false alarms since 6 out of the 60 buses that do not run on Main Street are blue. (In Bayesian terms, this branch is equivalent to p(blue|not Main), the conditional probability that a bus is blue given that it does not run on Main Street.)

One of the advantages of visualizing the blue bus problem using a frequency tree is that this method makes it easier to apply Bayesian reasoning to the problem. We only need two pieces of information—the number of hits and the number of false alarms—in order to find the Bayesian posterior (i.e., in order to determine the posterior probability that a bus is on Main Street given that it is blue).

In this particular example, we thus have 32 hits and 6 false alarms. In other words, there is an absolute total of 38 blue buses in town. Of these 38 blue buses, 32 buses run on Main Street (hits), while 6 do not (false alarms). Next, we find our Bayesian posterior by dividing the number of hits by the total number of hits and false alarms as follows:

$$p(Main|blue) = \frac{32}{32 + 6} = \frac{32}{48} = 0.84$$

Stated formally, there is an 84% posterior probability that a bus that runs down Main Street is blue. In plain language, since we know the accident occurred on Main Street, and since the hits outnumber the misses by a wide margin in this version of the problem, there is an extremely high posterior probability that the plaintiff's accident was caused by a blue bus. Thus, the plaintiff should pass the "preponderance of the evidence" test because it is more likely than not in this example that a blue bus caused the accident.

B. *Example #2 (Standard Format)*

Consider next what effect a change in the *prior probability* p(Main) has on the *posterior probability* p(Main|blue). In our original version of the blue bus problem in Part III.A above (Example #1), we assumed that p(Main) was equal to 40% or 0.4. What happens, however, when p(Main) is lowered to 0.1? That is, what if only 10% instead of 40% of all city buses run through Main Street?

In summary, because our hit and miss rates remain constant,[44] the evidence in support of plaintiff's hypothesis has not changed, and thus the degree or level of Bayesian updating in this example does not change. But because the *prior probability* is now much lower (10% instead of 40%), we would expect the *posterior probability* to also be lower, but the question is, how much lower?

---

44. For simplicity, we continue to use the conditional probabilities p(blue|Main) and p(blue|~Main) constant. Thus, the hit rate p(blue|Main) remains 0.8, and the miss rate p(blue|~Main)

To find out, we apply Bayesian reasoning to find the new posterior probability as follows:

1. Since 10% of city buses run on Main Street, and since 80% of those buses that travel down Main Street are blue, then only 8% of all city buses are blue and run through Main Street [0.1 * 0.8 = 0.08];

2. Since 90% of all city buses do not run down Main Street, and since only 10% of those buses are blue, then 9% of all city buses do not run through Main Street and are blue [0.9 * 0.1 = 0.09];

3. Lastly, since 0.06 + 0.09 = 0.17, a total of 17% of city buses are blue, and since 0.08 divided by 0.17 = 0.47, this means that there is only a 47% posterior probability that a certain bus runs on Main Street, *given that it is blue*.

Notice that the prior probability p(Main) has a substantial effect on the posterior probability p(Main|blue). Specifically, since so few buses run through Main Street (the scene of the accident), there is low probability that a blue bus caused the plaintiff's accident in this version of the blue bus problem, even when 80% of the buses that run on Main Street are blue. This finding, in turn, helps explain why the standard version of the blue bus problem (as summarized in Part II.C above) does not pass the "preponderance of the evidence" test: since the standard version of the blue bus case, such as Professor Tribe's blue bus hypothetical, provides no information about p(Main), the posterior probability p(Main|blue) can either be low (as in this example) or high (as in the next example).

---

also remains 0.1, as in Example #1. We hold these conditional probabilities constant in order to test what effect a change in the prior probability has on the posterior probability.

Example #2 (visual format)

We now present our second version of the blue bus problem visually using absolute numbers as follows:

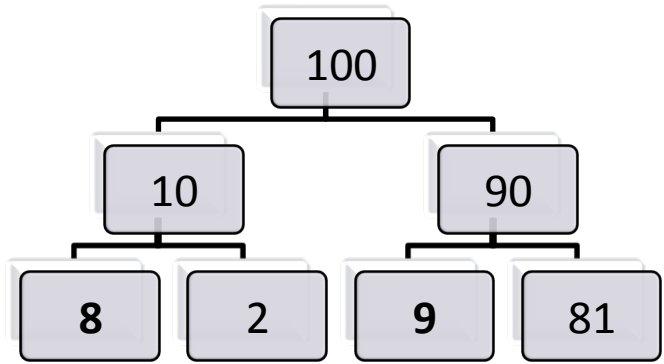

Again, we use absolute numbers instead of percentages or probabilities, starting with a population of 100 city buses.

First, recall our presentation of this particular example in the standard probability format. Although we ended up revising our prior probability *upward* from 0.1 to 0.47, we concluded that a Bayesian posterior of 0.47 does not pass the "preponderance of the evidence" test. Now, consider the absolute number of "hits" and "false alarms" on the bottom row of the frequency tree. We immediately notice that the 9 false alarms exceed the 8 hits by a small margin. This is an important observation. Without having to perform any complicated computation, we can immediately conclude that a random blue bus is slightly less likely to run on Main Street than a bus of another color, even though 80% of the buses that run on Main Street are blue buses.

The frequency tree above explains why. In brief, the small number of false alarms exceeds the number of hits because only 10 of the 100 city buses in this example run on Main Street, but 8 of these 10 buses are blue (since our hit rate in this example is 0.8, and 80% of 10 is 8), and of the 90 buses that do not run on Main Street, only 9 are blue. Thus there are more false alarms than hits.

### C. *Example #3 (Standard Format)*

Consider what happens when we increase the *prior probability* p(Main) that the buses in our population of city buses run on Main Street. In the previous example, we assumed that p(Main) was equal to 0.1. What happens, however, when p(Main) is raised to 0.8?[45] Since the prior probability is now much greater, we would also expect the posterior probability to be greater. In particular, we wish to discover how much greater will the posterior probability be when p(Main) is raised from 0.1 to 0.8,?

Again, for simplicity, we continue to hold the hit and miss rates constant, so the first conditional probability p(blue|Main) remains 0.8, and likewise, the second conditional probability p(blue|~Main) also remains 0.1. Given this information, we apply standard Bayesian reasoning to calculate the posterior probability as follows:

1. Since 80% of all city buses run on Main Street, and since 80% of those buses are blue, then 64% of all city buses are blue and run through Main Street [0.8 * 0.8 = 0.64];

2. Since 20% of all city buses do not run down Main Street, and since only 10% of those buses are blue, then 2% of all city buses do not run through Main Street and are blue [0.2 * 0.1 = 0.02];

3. Lastly, since 0.64 + 0.02 = 0.66, a total of 66% of city buses are blue, and since 0.64 divided by 0.66 = 0.96, this means that there is a whopping 96% posterior probability that a certain bus runs on Main Street, *given that it is blue*.

In other words, the prior probability has a substantial effect on the posterior probability, even when the hit and miss rates remain the same. Specifically, since most buses run through Main Street (since the value for p(Main) is high in this case), and, since we know the accident occurred on Main Street, there is an extremely high posterior probability in this version of the blue bus problem that the plaintiff's accident was caused by a blue bus. Thus, the plaintiff in this example should pass the "preponderance of the evidence" test because it is more likely than not that a blue bus caused the accident.

---

45. That is, suppose 80% instead of only 10% of all city buses run through Main Street.

Example #3 (visual format)

We now present this version of the blue bus problem visually as follows, using absolute numbers instead of percentages:

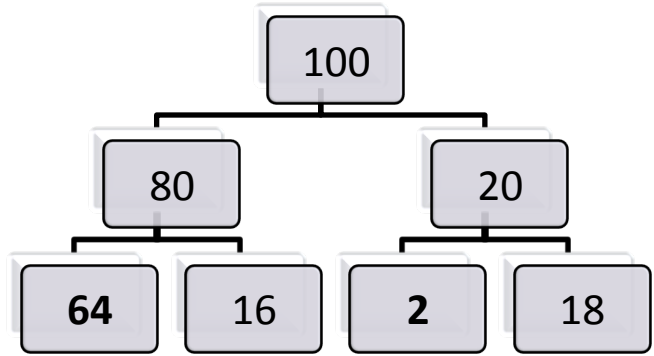

Without having to perform any complicated computations, we immediately notice 64 "hits" compared with just 2 "false alarms." In Bayesian terms, a random blue bus in this example is far more likely to run on Main Street than a bus of any other color because the number of hits exceeds the number of false alarms by a wide margin.

Again, the frequency tree explains this conclusion in three steps: most of the city buses run on Main Street (80 out of 100) (step one), and most of these buses in turn are blue (64 out of 80) (step two), while only 2 of the 20 buses that do not run on Main Street are blue (step 3). Thus, because there are far more hits than false alarms, it is more likely than not (in terms of the preponderance of the evidence standard) that the bus that caused the plaintiff's accident was a blue bus. The plaintiff in this particular example should thus pass the "preponderance of the evidence" test.

*D. Example #4 (Standard Format)*

Next, consider what happens when the hit rate, p(blue|Main), is greater than the miss rate, p(blue|~Main). Expressed in the standard Bayesian probability format, what happens when we return the prior probability back to its original value, 40%, and instead increase the first conditional

probability or hit rate to 95%? In this version of the blue bus problem, our revised priors are as follows:

- The probability that a city bus runs on Main Street is 0.4;
- The probability that a certain bus is blue, given that it runs on Main Street, is now 0.95;
- The probability that a certain bus is blue, given that it does not run on Main Street, is still 0.1.

Given this revised data, we apply Bayesian reasoning to find the new posterior probability as follows:

1. Since 40% of all city buses run on Main Street, and since 95% of those buses are blue, then 38% of all city buses are blue and run through Main Street [0.4 * 0.95 = 0.38];

2. Since 60% of all city buses do not run down Main Street, and since only 10% of those buses are blue, then 6% of all city buses do not run through Main Street and are blue [0.6 * 0.1 = 0.06];

3. Since 0.38 + 0.06 = 0.44, a total of 44% of city buses are blue, and since 0.38 divided by 0.44 = 0.86, this means that there is a whopping 86% posterior probability that a certain bus runs on Main Street, *given that it is blue*.

Notice that the first conditional probability p(blue|Main) in this example is far larger than the other conditional probability p(blue|~Main). In plain language, this means that there are significantly more buses that are blue and that run on Main Street than buses that are blue but do not run on Main Street. Based on the evidence in this example, we must update our posterior probability that a bus is on Main Street given that it is blue upwards—in this example, from 40% percent up to 86%. Although the plaintiff in the blue bus problem may not have seen the color of the bus that caused her to swerve, she knows that the accident occurred on Main Street, and the Bayesian updating in this example strongly supports her hypothesis that a blue bus caused her to swerve.

Also, before proceeding, notice that this version of the blue bus problem corresponds most closely to the facts in *Smith v. Rapid Transit*, discussed in Part II.B of this article. In essence, the plaintiff's evidence in *Smith* showed that only the equivalent of blue buses were authorized to run on Main Street, but as the court noted, the plaintiff's probabilistic proof "did not preclude private or chartered buses from using this street."[46] According to the court's reasoning, there was always a small probability that a non-blue bus was travelling down Main Street at the time of the accident. What the court should have been asking, however, was "how small" was this small probability?

Example #4 (visual format)

Using our frequency tree format, we now present this example visually as follows:

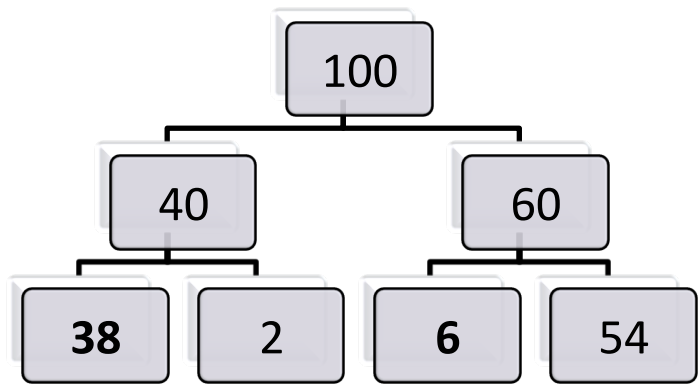

Again, without having to perform any complicated computations, we immediately notice 38 "hits" compared with just 6 "false alarms." Thus, as in the previous example, a random blue bus in this particular example is more likely to run on Main Street than a bus of any other color because the number of hits exceeds the number of false alarms by a wide margin. In legal terms, the plaintiff in this example should pass the "preponderance of the evidence" test.

E. *Example #5 (Standard Format)*

Now, consider what happens when both the hit rate and miss rate are high. For example, what happens when the hit rate p(blue|Main) is 95% and the miss rate p(blue|~Main) is 80%, while the base rate p(Main) is

---

46. *Smith*, 58 N.E.2d at 755.

constant at 40%, as in our original example?[47] Our revised priors are now as follows:

- The probability that a city bus runs on Main Street is still 0.4;

- The probability that a certain bus is blue, given that it runs on Main Street, is 0.95;

- The probability that a certain bus is blue, given that it does not run on Main Street, *is now 0.8.*

Given these revised pieces of information, we apply standard Bayesian reasoning to update our priors as follows:

1. Since 40% of all city buses run on Main Street, and since 95% of those buses are blue, then 38% of all city buses are blue and run through Main Street [0.4 * 0.95 = 0.38];

2. Since 60% of all city buses do not run down Main Street, and since 80% of those buses are blue, then 48% of all city buses do not run through Main Street and are blue [0.6 * 0.8 = 0.48];

3. Since 0.38 + 0.48 = 0.86, a total of 86% of city buses are blue;

4. Lastly, since 0.38 divided by 0.86 = 0.44, there is only a 44% posterior probability that a certain bus runs on Main Street, *given that it is blue*.

In summary, notice that the prior probability, p(Main), in this example and in the previous example did not change. In both cases, p(Main) was set at 40%. Notice too that the posterior probability in this example, 44%, remained close to our prior probability, 40%. In the previous example, by contrast, the posterior probability increased to 86%, a large departure from our original prior probability.

Thus, by comparing this example to the previous one, we see how the difference between the two conditional probabilities (the difference between the rate of hits and the rate of misses) determines the level of updating in our Bayesian model. In sum, if the difference between the two conditional probabilities is small (as in this example), then the effect on the prior probability is small. We do not need to update or revise our prior probability too greatly. If, however, the difference between the two conditional probabilities is large (as in the previous example), then this difference will lead us to revise our prior probability by a large amount as well.

---

47. We continue to hold p(Main) constant in order to test the effect of changes in the hit and miss rates.

Example #5 (visual format)

We now present this same example visually as follows:

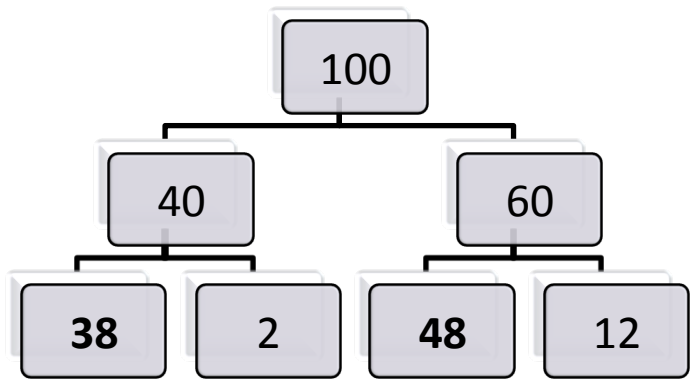

Once again, without having to perform any complicated computations, we immediately notice the total or absolute number of hits and false alarms: 38 hits versus 48 false alarms. Since there are more false alarms than hits, this means that a random blue bus is less likely to run on Main Street than a bus of another color. Given this observation, the plaintiff should not pass the "preponderance of the evidence" test.

F. *Example #6 (Standard Format)*

Next, consider what happens when the miss rate is greater than the hit rate. Expressed in Bayesian terms, what happens when the second conditional probability is larger that the first conditional probability? For example, what if p(blue|Main) is only 30%, while p(blue|~Main) is 60%?[48] Our revised priors can thus be stated formally as follows:

- The probability that a city bus runs on Main Street is 0.4;

---

48. Again, we hold p(Main) constant at 40% in order to test the effect of changes in the hit and miss rates.

- The probability that a certain bus is blue, given that it runs on Main Street, is 0.3;
- The probability that a certain bus is blue, given that it does not run on Main Street, is 0.6.

Stated formally, the second conditional probability ("the probability that a bus is blue given that it does not run on Main Street") is twice as large as the first conditional probability ("the probability that a bus is blue given that it does run on Main Street"). Given this new data, we apply standard Bayesian reasoning as follows:

- Since 40% of all city buses run on Main Street, and since 30% of those buses are blue, then 12% of all city buses are blue and run through Main Street [0.4 * 0.3 = 0.12];

- Since 60% of all city buses do not run down Main Street, and since 60% of those buses are blue, then 42% of all city buses do not run through Main Street and are blue [0.6 * 0.6 = 0.36];

- Since 0.12 + 0.36 = 0.48, a total of 48% of city buses are blue, and since 0.12 divided by 0.48 = 0.25, there is only a 25% posterior probability that a certain bus runs on Main Street, *given that it is blue*.

In other words, when the second conditional probability is larger than the first conditional probability, (i.e., when the rate of misses is larger than the rate of hits) we must update our prior probability *downward*—in this example, from 40% down to 25%.

Now, consider this result in light of the blue bus problem. When the rate of misses p(blue|~Main) is greater than the rate of hits p(blue|Main), this means that there are more buses that are blue and do not run on Main Street than buses that are blue and do run on Main Street. Since our hypothesis is that a blue bus caused the plaintiff's accident, our conditional probabilities are telling us that this hypothesis is more likely to be false because it is more likely for a bus on Main Street not to be blue.

Example #6 (visual format)

We now present this example visually as follows:

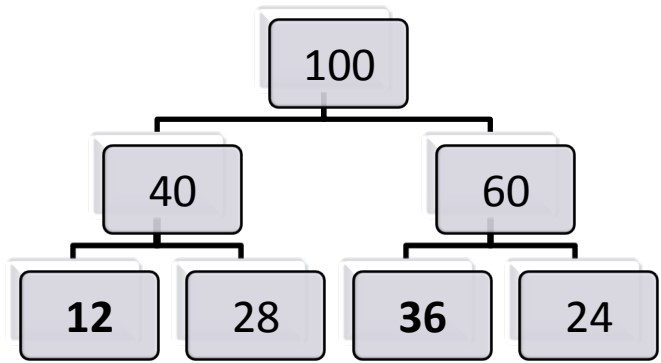

Without having to perform any complicated computations, we immediately notice 12 hits and 36 false alarms. As in the previous example, there are more false alarms than hits. In Bayesian terms, a random blue bus is less likely to run on Main Street than a bus of another color, and given this observation, the plaintiff in this example should not pass the "preponderance of the evidence" test.

G. *Example #7 (Standard Format)*

Lastly, suppose the rate of hits and false alarms are the same. In Bayesian terms, what if our conditional probabilities are the same? For example, our hit rate p(blue|Main) is now 0.8 and our miss rate p(blue|~Main) is now 0.8, while p(Main) is constant at 0.4. Our revised priors are now as follows:

- The probability that a city bus runs on Main Street is 0.4;

- The probability that a certain bus is blue, given that it runs on Main Street, is 0.8;

- The probability that a certain bus is blue, given that it does not run on Main Street, is 0.8.

Given this new information, we again apply the standard Bayesian method as follows:

1. Since 40% of all city buses run on Main Street, and since 80% of those buses are blue, then 32% of all city buses are blue and run through Main Street [0.4 * 0.8 = 0.32];

2. Since 60% of all city buses do not run down Main Street, and since 80% of those buses are blue, then 48% of all city buses do not run through Main Street and are blue [0.6 * 0.8 = 0.48];

3. Since 0.32 + 0.48 = 0.80, a total of 80% of city buses are blue;

4. Lastly, since 0.32 divided by 0.80 = 0.4, there is a 40% posterior probability that a certain bus runs on Main Street, *given that it is blue*.

Notice that the posterior probability in this example is the same as the prior probability. Furthermore, this will always be the case when the hit and miss rates are the same. In general, when both conditional probabilities are the same, the conditional probabilities end up exerting the same amount of force on our Bayesian update—meaning that our Bayesian update provides us with no new information from which to revise our original prior probability.

In terms of the blue bus problem, if a bus is just as likely to be blue given that it runs on Main Street (the hit rate or first conditional probability) as it is likely to be blue given that it does not run on Main Street (the miss rate or second conditional probability), the fact that the defendant bus line owns all the blue buses in town does not provide any new information about whose bus caused the plaintiff's accident. Thus, with no new information, we are unable to update or revise our prior probability.

Example #7 (visual format)

We now present this example visually as follows:

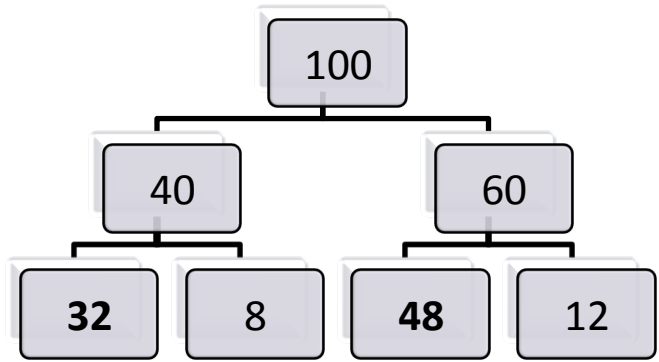

As in the previous two examples, there are more false alarms than hits: 48 false alarms versus 32 hits. In Bayesian terms, since there are more false alarms than hits, a random blue bus is less likely to run on Main Street than a bus of another color. Given this observation, the plaintiff should not pass the "preponderance of the evidence" test. The larger lesson here is that defendants, not just plaintiffs, can potentially use Bayesian methods.

Thus far, we have used frequency trees and frequency grids to visualize our Bayesian models. Next, we present an alternative visualization of our Bayesian approach to the blue bus problem.

*H. An Alternative Visualization of the Bayesian Approach*

Here, we present an alternative visualization of the blue bus problem based on the work of Luke Muehlhauser and Eliezer Yudkowsky.[49] Consider the following diagram:

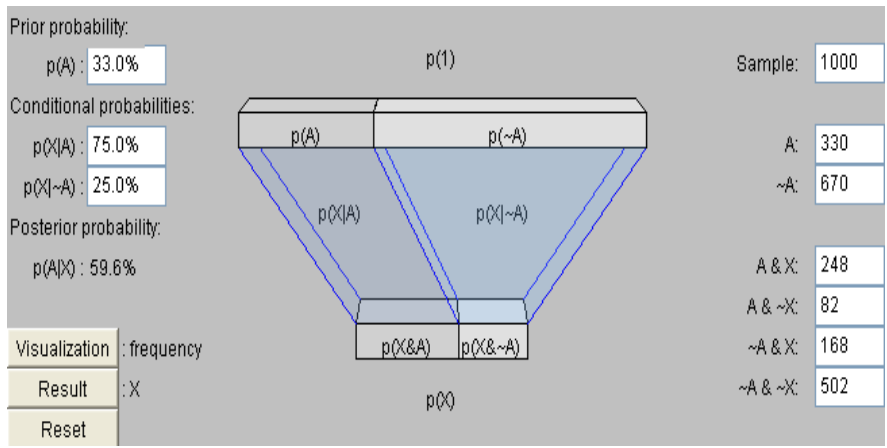

The diagram conveys the following information:

- The bar on the top of this diagram is divided between p(A) and p(not A), where "A" is the condition of being on Main Street. In brief, this bar shows the prior probability that a bus (of any color) runs on Main Street. Since the probability in this hypothetical example is 33%, the division between the two sides of the bar is just left of center.[50]

- The slanted line in the middle of the diagram represents the effect of hits and misses, or the effect of the conditional probabilities, on the prior probability, where "X" is the condition of being blue. The first conditional probability, the hit rate, is p(X|A) (i.e., "the probability that a bus is blue given that it runs on Main Street"). Similarly, the second conditional

---

49. *See* Luke Muehlhauser, *An Intuitive Explanation of Eliezer Yudkowsky's Intuitive Explanation of Bayes' Theorem* (Dec. 18, 2010), http://commonsenseatheism.com/?p=13156 (last visited on Oct. 12, 2014). *See also* Eliezer S. Yudkowsky, *An Intuitive Explanation of Bayesian Reasoning* (June 4, 2006), available at http://yudkowsky.net/rational/bayes (last visited on Sept. 26, 2014). The Bayesian diagram below appears in Yudkowsky.

50. If the prior probability in our example had been 50% instead of 40%, the division between the two sides would have appeared in the center.

probability is p(X|~A), or "the probability that a bus is blue given that it does not run on Main Street." This is the rate of misses. Since the hit rate p(X|A) is greater than the miss rate p(X|~A) in this example, the middle line is slanted from left to right, and the degree of this slant reflects the probability regarding the color of a particular bus running on Main Street.

- Finally, the bottom row represents the revised or posterior probability that a bus is blue and runs on Main Street, or p(X & A), as well as the posterior probability that a bus is blue and does not run on Main Street, p(X & ~A).

Also, before proceeding, notice that the bar at the top of the diagram is much wider than the bar at the bottom. This differential occurs because the top bar refers to the collection of all buses that run or do not run on Main Street, while the bottom bar refers only to the collection of blue buses that run or do not run on Main Street.[51]

Now, returning to the diagram, notice the differential in *proportions* between the left side of the top bar and the left side of the bottom bar. This differential tells us that the proportion of buses that are blue and that run on Main Street (the left side of the bottom bar) is somewhat larger than the proportion of all buses that run on Main Street (the left side of the top bar). Thus, based on the evidence in this example, the hit and miss rates p(X|A) and P(X|~A), we must update our posterior probability that a bus is on Main Street given that it is blue upwards—in this example, from 33% percent up to 59.6%. Although the plaintiff in the blue bus problem may not have seen the color of the bus that caused her to swerve, she knows that the accident occurred on Main Street, and the Bayesian updating in this example strongly supports her hypothesis that a blue bus caused her to swerve.

Does this particular example pass the "preponderance of the evidence" test? This example ought to pass the preponderance test because in this case it is more likely than not that a blue bus caused the accident.

IV. THE CASE FOR BAYES

Thus far, we have restated the famous blue bus thought-experiment and have presented our alternative Bayesian visualization of the blue bus problem. But what about the general questions we posed at the beginning of this paper regarding the role of mathematics and mathematical reasoning in law?[52] Specifically, what larger lessons can we hope to learn from our simple Bayesian model of the blue bus problem?

---

51. *See generally* Muehlhauser, *supra* note 49.

Recall the legal issue in *Smith v. Rapid Transit* and the question posed by the blue bus thought-experiment itself: the problem of probabilistic proof in legal trials. Specifically, what role should probabilistic proof play in the legal process? What role should the Bayesian methods proposed in this paper and by others play in adjudication?[53] I shall conclude this paper by summarizing and discussing the main virtues of Bayesian reasoning generally and of my Bayesian approach to evidence.[54]

*A. What Bayes is Not*

To avoid confusion, we begin by stating what Bayesian reasoning is *not*. It is not a magic conjurer, Delphic oracle, or "veritable sorcerer."[55] Bayesian reasoning is not a method for eliminating uncertainty; it is a method for measuring our level of uncertainty.[56] Put another way, the Bayesian approach to evidence is not only a method of drawing inferences, but it is also a method of measuring the strength or weakness of such inferences. In the legal process context, the Bayesian approach is not a method for deciding the ultimate guilt or innocence of a defendant. Instead, it is a method for measuring or evaluating the strength of a party's evidence of guilt or innocence, and thus Bayesian methods may not only be used *offensively* by a moving party (plaintiff or prosecutor) to measure the strength of his case; such methods can also be used *defensively* by a defendant to test or challenge the strength of the moving party's case.

---

52. It is worth noting that these same questions were first raised two centuries ago by such intellectual giants as Pierre Simon Laplace and the Marquis de Condorcet. *See generally*, Bertsch McGrayne, *supra* note 1, at 27–28. Here, our focus is on the problem of probabilistic proof.
53. It is worth noting that Tribe's original critique of mathematical proof in law (*see* Tribe, *supra* note 4) was a response to a paper proposing Bayesian methods in law.
54. For a more systematic explanation of Bayesian reasoning, *see generally* Yudkowsky, *supra* note 49. *See also* Colin Howson & Peter Urbach, *Bayesian Reasoning in Science*, 350 NATURE 371 (1991).
55. *Collins*, 438 P.2d at 33.
56. *See, e.g.*, Howson & Urbach, *supra* note 54, at 372 (Bayesian reasoning is a method of "characterizing a scientific conclusion about a hypothesis as a statement of its probability"); Fienberg & Schervish, *supra* note 7, at 773 ("Bayesian probability theory . . . provide[s] both a framework for quantifying uncertainty and methods for revising uncertainty measures in the light of acquired evidence.").

*B. A Legal Trial is Just a Test*

We often assume that trials are a search for the truth.[57] Yet, truth itself is a probabilistic ideal, not an absolute one, since there might be different versions or interpretations of the truth. Moreover, from the perspective of a trial attorney and the litigating parties, a trial is a risk-taking activity, a game whose main object is to "win," regardless of truth. Metaphorically speaking, then, a legal trial is more like poker and less like science.[58]

Of course, it helps to have the truth on one's side, but my point here is that having the truth on one's side is neither a necessary nor sufficient condition for winning at trial (the ultimate goal of litigation from the perspective of the parties). Truth, too, is a probabilistic concept, for there can be competing conceptions of the truth in any given case.[59] In law, this means that a legal trial is not a dispassionate or scientific search for truth. Rather, a legal trial is a bet or wager on which party's story or version of the truth is more likely to be accepted as true.

In sum, instead of a search for truth, a legal trial is a test—a test of the evidence one offers in support of one's version of the truth. Thus, a legal trial is more like a medical test or a spam filter, but instead of testing for cancer, HIV, or spam, litigation simply tests the strength of the moving party's case. In Anglo-American law, the moving party is either the State (e.g., the prosecution in a criminal case) or a private party (e.g., the plaintiff in a civil case), and when a prosecutor or plaintiff goes to trial, he is literally putting his allegations (his case) to the test. In either case, civil or criminal, the moving party must submit sufficient evidence to pass the relevant test. Sometimes the test is pass/fail (e.g., the preponderance of the evidence standard); sometimes the test requires a higher standard of evidence (e.g., the reasonable doubt standard). In either case, the question is the same: has the prosecution proven its case beyond a reasonable doubt? Or, has the plaintiff proven his case by a preponderance of the evidence? This, then, is what is meant when describing a trial as a test.

---

57. *See, e.g.*, Cheng, *supra* note 7, at 1273 ("the legal system's purpose is to discover a single, stable 'truth' or narrative"). *See also* Thomas A. Mauet, *Trial Techniques*, 8th edition, New York: Aspen (2010), p. 24 ("A theory of the case is a clear, simple story of 'what really happened' from your point of view . . . Trials are in large part a contest to see which party's version of 'what really happened' the jury will accept as more probably true"). In the criminal context, *see Brady v. Maryland*, 373 U.S. 83, 86 (1963) (citation omitted). For an extended discussion of the *Brady* rule in criminal cases, *see* Robert Hochman, Brady v. Maryland *and the Search for Truth in Criminal Trials*, 63 U. CHI. L. REV. 1673 (1996).
58. For a simple poker model of litigation, see Enrique Guerra-Pujol, *The Poker-Litigation Game* (Working Paper Series, Dec. 26, 2012), *available at* http://papers.ssrn.com/sol3/papers.cfm?abstract_id=2193993. For a risk-based analysis of litigation, see Tristan Barnett, *Applying the Kelly Criterion to Lawsuits*, 9 LAW, PROB. & RISK 139 (2010).
59. *Cf.* Cheng, *supra* note 7, at 1259 ("the preponderance standard [in civil cases] is better characterized as a probability ratio, in which the probability of the plaintiff's story of the case is compared with the defendant's story of the case").

### C. A Test is Separate from Reality

When we use a medical test, or a spam test, or a legal test, we must cognitively distinguish that the test is separate from the underlying hypothesis or event being tested. Put another way, "tests are not the event."[60] A spam filter tests for spam, but the spam test itself is separate from the event of actually having a spam message, and by the same token, a test for cancer or HIV is separate from the event of actually having cancer or HIV. Or, in the words of one Bayesian:

> Even a *useful* mammography test does not actually change [the underlying reality] *whether or not a woman has cancer*. She either has cancer or she doesn't. Reality is not uncertain about whether or not the woman has cancer. *We* are uncertain about whether or not she has cancer. It is our information, our judgment, that is uncertain, not reality itself.[61]

Likewise, a legal trial or "legal test" is separate from the underlying condition being tested: a defendant's guilt or innocence. Again, the defendant either has or has not committed a wrongful act (e.g., a crime, a tort, a breach of contract, etc.).[62] Reality itself is not uncertain about the defendant's conduct. It is we (the jury or the trier or fact) who are uncertain about the defendant's guilt or innocence.[63]

As an aside, this point helps explain the logic of our Bayesian approach to adjudication. A Bayesian or any probabilistic method for trying cases is not designed to tell us with 100% certainty whether a given defendant is guilty or innocent. No method or test could produce such perfect outcomes. Instead, our Bayesian approach is simply an alternative method for testing the strength of the moving party's case, that is, for testing how likely the plaintiff or prosecutor has proven his or her case.

---

60. This phrase is from Kalid Azad, *An Intuitive (and Short) Explanation of Bayes' Theorem* (May 7, 2007), *available at* http://betterexplained.com/articles/an-intuitive-and-short-explanation-of-bayes-theorem/ (last visited on Sept. 26, 2014).
61. Muehlhauser, *supra* note 49, at 8 in the pdf format, italics in original (explaining the ideas of Yudkowsky, *supra* note 49).
62. As our editors have pointed out, the legal elements of a wrongful act—that is, what constitutes a particular crime or tort—are human-constructed elements, since we ultimately define what elements constitute a tort or crime, such as "battery," for example. As a result, the underlying metaphysical reality (e.g., "did this defendant commit a battery?") might not congruently align with the human-defined elements of the wrongful act. In this case, stating that "a defendant either has or has not committed a crime" would present a false dichotomy. Nevertheless, there is no reason in principle why we could not apply Bayesian methods to the task of interpreting what behavior constitutes a wrongful act in the first place. That is, we could apply Bayesian methods to predict what interpretation of law is most likely to prevail in a given close case.
63. Even the practice of science, like litigation, is just a series of tests. There is a test for a given phenomenon, and there is the event of the phenomenon itself. *See* Azad, *supra* note 60.

*D. All Tests are Imperfect*

Tests sometimes detect things that do not exist (false positives or "false alarms") and miss things that do exist (false negatives or "misses").[64] Consider spam filters, which are designed to detect spam or unwanted email messages. The problem is that no spam filter is perfect. Sometimes the spam net is cast too wide and some non-spam email messages fall into the spam filter and into one's junk email folder; sometimes the spam net is not wide enough and spam messages get past the filter and into your regular inbox.

By analogy, the same problem occurs in litigation. A "frivolous" civil case, or criminal case in which the prosecutor has "overcharged" the defendant, can be compared to spam. In a perfect legal system, judges and juries would be able to detect and distinguish frivolous claims from valid ones (in civil cases) and such superfluous criminal charges from substantial ones (in criminal cases). But judges and juries make mistakes. We all do. Sometimes judges dismiss valid claims (misses) and allow frivolous claims to go to trial (false alarms). Likewise, sometimes juries convict innocent men (false alarms) and allow the guilty to go free (misses).

*E. Why Bayes is Useful: It Puts Tests to the Test*

Bayesian methods are useful because they allow one to update and test one's prior probability of a given event (e.g., x has cancer, y is a spam message, z committed a crime or a tort). In the case of a legal trial, the Bayesian approach is a formal method for updating one's prior beliefs about a defendant's guilt or innocence. Put another way, Bayesian reasoning is a method for testing the reliability and accuracy of a legal trial.[65] It is a way of testing legal tests, a test of a test. This principle, the

---

    64. As we noted in Part III above, the terms "false alarms" and "misses" were used by Engel and Gigerenzer. Engel & Gigerenzer, *supra* note 40, at 10.
    65. *See, e.g.*, Enrique Guerra-Pujol, *A Bayesian Model of the Litigation Game*, 4 EUR. J. LEGAL

idea that Bayes puts tests to the test, we believe is the most important one.

## V. CONCLUSION

This article represents a small first step in solving the problem of probabilistic proof in law. By focusing on a simple proof problem, the famous blue bus hypothetical, and by visualizing this problem using Bayesian methods, we have demonstrated the following lessons:

- Probabilistic proof—like all forms of proof—can in principle be subjected to Bayesian testing given sufficient information about hits and misses.

- Our simplified Bayesian testing procedure consists of comparing the number of hits and misses.

- Our testing procedure measures the strength of the evidence being tested.

- Visual presentation of a given evidence problem can be useful because it can simplify the problem.

- But, to apply our test protocol, we need information about base, hit, and miss rates.

In sum, legal scholars, lawyers, and judges should reconsider the potential usefulness of Bayesian methods at trial to test and evaluate the strength of probabilistic proof, instead of rejecting such proof altogether. To the extent law is nothing more than a prediction of what courts, juries, and legislators will do,[66] the law of the future should be open to Bayesian methods.

---

STUDIES 220 (2011).
    66. *See* O.W. Holmes, *The Path of the Law*, 10 HARV. L. REV. 457, 460–61 (1897) ("The prophecies of what the courts will do in fact, and nothing more pretentious, are what I mean by the law.").